\documentclass[sigchi]{acmart}

\AtBeginDocument{%
  \providecommand\BibTeX{{%
    \normalfont B\kern-0.5em{\scshape i\kern-0.25em b}\kern-0.8em\TeX}}}

\setcopyright{acmcopyright}
\copyrightyear{2023}
\acmYear{2023}
\acmDOI{https://doi.org/10.1145/3593013.3594002}

\acmJournal{TOG}
\acmVolume{37}
\acmNumber{4}
\acmArticle{111}
\acmMonth{8}

\acmConference[FAccT ’23]{ACM Conference}{June 12-15, 2023}{Chicago, USA}

\copyrightyear{2023}
\acmYear{2023}
\setcopyright{rightsretained}
\acmConference[FAccT '23]{2023 ACM Conference on Fairness, Accountability, and Transparency}{June 12--15, 2023}{Chicago, IL, USA}
\acmBooktitle{2023 ACM Conference on Fairness, Accountability, and Transparency (FAccT '23), June 12--15, 2023, Chicago, IL, USA}
\acmDOI{10.1145/3593013.3594002}
\acmISBN{979-8-4007-0192-4/23/06}

\title{Stronger Together: on the Articulation of Ethical Charters, Legal Tools, and Technical Documentation in ML}

\author{Giada Pistilli}
\orcid{0000-0003-4941-0505}
\affiliation{%
  \institution{Hugging Face}
  \country{France}}
\email{giada@huggingface.co}

\author{Carlos Muñoz Ferrandis}
\orcid{0000-0001-7178-762X}
\affiliation{%
  \institution{Hugging Face}
  \country{Spain}}
\email{carlos@huggingface.co}

\author{Yacine Jernite}
\orcid{0000-0002-8053-6862}
\affiliation{%
  \institution{Hugging Face}
  \country{United States}}
\email{yacine@huggingface.co}

\author{Margaret Mitchell}
\orcid{0000-0001-7043-6545}
\affiliation{%
  \institution{Hugging Face}
  \country{United States}}
\email{meg@huggingface.co}

\begin{abstract}
The growing need for accountability of the people behind AI systems can be addressed by leveraging processes in three fields of study: ethics, law, and computer science.
While these fields are often considered in isolation, they rely on complementary notions in their interpretation and implementation.
In this work, we detail this interdependence and motivate the necessary role of collaborative governance tools in shaping a positive evolution of AI.
We first contrast notions of compliance in the ethical, legal, and technical fields; we outline both their differences and where they complement each other, with a particular focus on the roles of ethical charters, licenses, and technical documentation in these interactions.
We then focus on the role of values in articulating the synergies between the fields and outline specific mechanisms of interaction between them in practice.
We identify how these mechanisms have played out in several open governance fora: an open collaborative workshop, a responsible licensing initiative, and a proposed regulatory framework. 
By leveraging complementary notions of compliance in these three domains, we can create a more comprehensive framework for governing AI systems that jointly takes into account their technical capabilities, their impact on society, and how technical specifications can inform relevant regulations. Our analysis thus underlines the necessity of joint consideration of the ethical, legal, and technical in AI ethics frameworks to be used on a larger scale to govern AI systems and how the thinking in each of these areas can inform the others. % We conclude by encouraging similar exploratory research of the three compliances' articulation in practice.
\end{abstract}

\ccsdesc[500]{Social and professional topics}

\keywords{AI Governance, Applied Ethics, ML Licensing, Documentation, AI Policy}

\begin{document}

\maketitle

\section{Introduction}

As AI systems \footnote{In this paper, we make the distinction between an Artificial Intelligence (AI) system and a Machine Learning (ML) artifact: the former is a fully deployed system that relies on AI (e.g., a resume screening software); the latter is any stand-alone object that has to do with ML (e.g., ML models).} have been taking a growing place in technological developments of recent years, elaborating mechanisms to govern these systems and shape their evolution in ways that most benefit a diverse range of stakeholders with different priorities and levels of access to their development has become a necessity.

One notable focus of recent efforts to that end has been the design of numerous guiding principles for AI systems, formalized in ethical charters by governments, civil society, national and international institutions, research laboratories, and other types of organizations~\cite{jobin2019global,UNESCO_2021}. Their purpose is twofold: on the one hand, they seek to frame the development of AI systems \cite{vandepoel2016ethical} and, on the other, to guide their proper use \cite{hine2022artificial}, all in order to protect the affected human stakeholders.
However, notwithstanding their widespread use in medical ethics \cite{campbell1997medical}, ethical charters are still a long way away from supporting the agile operationalization of ethical principles that would make them effective ethical instruments\cite{IEEEGlobalInitiative}.
Work on principles has also been accompanied by regulatory efforts
\cite{europa2021proposal}
to start extending existing legislation in a way that better accounts for the new technical reality 
\cite{europa2022liability},
as well as more technically focused proposals to better document and specify the working of the systems at play~\cite{GebruMorgensternEtAl2020,MitchellWuEtAl2019}.%\\

%While navigating the complex relationship between legal and ethical compliance, the former might sometimes intersect with the latter, as both phenomena are not mutually exclusive nor inherently articulated.
Ethical and legal notions of compliance can intersect in various ways and are neither mutually exclusive nor inherently articulated.
For instance, in order for a corporate director to be morally compliant with a company's code of ethics that features openness and transparency as core values, they would have to follow corporate, financial, or banking law-specific provisions outlining internal duties for disclosure of information to managers. However, this value of transparency also entails good communication practices more broadly than what is strictly required for legal compliance.
Notwithstanding the interrelations between ethical and legal compliance, legal compliance does not inherently entail moral compliance.

A similar example of this complex relationship can be found in the ongoing debate over the legality of data scraping techniques employed in training generative AI systems~\cite{krotov2018legality}, with consent playing a pivotal role.
The value of consent is often regarded as a cornerstone of ethical frameworks, emphasizing respect for individual autonomy and data privacy.
Even when an interpretation of fair use in copyright laws, such as those under US copyright law, permits data scraping for commercial objectives, the practice may still be considered unethical if it disregards the element of consent.
Engaging in the non-consensual use of copyrighted images for large-scale machine learning training can potentially be legally compliant while simultaneously being viewed as immoral by specific art communities, collectives, or individual creators who place a strong emphasis on respecting consent and safeguarding their artistic works~\cite{glaze}.

An additional source of complexity when assessing compliance comes from its dependence on understanding the specific technical behaviors of AI systems. For example, whether deploying a language model violates the privacy of its training dataset's data subject will depend on the model's ability and likelihood of memorizing specific documents~\citep{Carlini2022QuantifyingMA}, and metrics quantifying biases in a system can help demonstrate how systems might run afoul of anti-discrimination laws~\citep{Buolamwini2018GenderSI}. This gives technical documentation a dual role in enabling compliance and in informing ethical and legal frameworks.

% Meanwhile, it is crucial to recognize that adhering to regulations through technical choices frequently necessitates making value-based trade-offs, such as balancing safety against diversity or privacy against transparency, which can significantly impact various stakeholders.%\\

%In this article, we describe how the effectiveness of these governance artifacts increases when a framework of coordination and complementarity is set up from the beginning of the development of a particular AI system.
%In this context, reality often collides with theory when putting certain guiding principles or moral values into practice.
%Moreover, the different scientific fields that should naturally come together in consolidating what can be called interdisciplinary AI ethics often conflict when confronted with the concrete applications of AI systems.
% For example, it can be challenging to distinguish between the consequences of moral or legal responsibility; moreover, ethical compliance, even for internationally deployed systems, is often mistaken as fulfilling a particular territory's national regulatory requirements. 

In this work, we aim to shed light on specific mechanisms of interaction between the ethical, legal, and technical aspects of governance of AI systems to inform an analysis of their synergies, complementary aspects, and the role of joint consideration of these three fields in strengthening their ability to shape the development of the technology.
We review recent work on sociotechnical considerations of AI, as well of new categories of ethical, legal, and technical artifacts aimed at supporting its governance, in Section~\ref{sec:background}.
Section~\ref{sec:compliances} then reviews three definitions of compliance corresponding to the three fields of study considered to outline their similarities, differences, and when they need to rely on each other to function.
Section~\ref{sec:applications} describes three case studies at the intersection of two or more of these domains in AI governance and development: an open research collaboration focused on developing a Large Language Model, a new licensing paradigm for ML artifacts, and the role of model cards in the upcoming EU AI Act.
Section~\ref{sec:discussion} then illustrates commonalities in these intersections, and we conclude with a discussion of learnings and future directions of research in Section~\ref{sec:conclusion}.

\section{Background}
\label{sec:background}

% \paragraph{Related work}

Our analysis framework is set in the sociotechnical context of the exponential development of AI systems. Integrating social and technical elements in sociotechnical systems requires a comprehensive understanding of both their human and artificial components, as their effectiveness depends on how well they interact within a social, organizational, or legal context. This context is shaped by society's values, beliefs, norms, and policies \cite{Jones_2013}.

On this basis, we respond to Luciano Floridi’s call when he insists on interdisciplinarity in ethics when applied to the digital world \cite{Floridi2018SoftEA}. In Floridi's governance framework between soft and hard ethics, the latter has the ability to influence national and international digital governance regulations, making it a critical piece of communication between ethics, policy, and law. Within this frame of reference, we go beyond what Floridi suggests. In light of the need for accountability, specifically in developing AI systems, we propose a framework for analysis that incorporates computer science within Floridi's overview. We argue that, thanks to this technical component, ethics is able to conduct its testing and operationalize its values.\\

% From a more theoretical perspective, w
We base our interdisciplinary articulation work on the philosophy of law. In its theory, a close connection links philosophy, namely, ethics, and law. In philosophy, two schools of thought oppose each other: positivists think that law influences the intrinsic values of a given society \cite{Hart1961-HARTCO-9}, while other philosophers argue precisely the opposite \cite{dworkin_justice_2011}. According to Dworkin~\citep{dworkin_justice_2011}, ethics not only plays a vital role in shaping the nature and interpretation of the law, but it has the power to influence its interpretation and application. Following his reasoning, the law is a system of principles that reflects a society's values and beliefs, rather than a simple set of rules issued by an institution with legislative powers. Thus, the law becomes a coherent system of principles justified by their consistency with one another and with the broader values and beliefs of the community in which they apply. In this context, the law is endowed with an organic nature that constantly adapts and evolves according to new social situations \cite{dworkin_taking_1977}.

In this evolving context, tools or processes that can translate ethical values into concrete actions are often missing in the industrial AI development context. However, a few advances have been made in this regard. This includes auditing frameworks such as Raji et al. \cite{Raji2020} that guide how to structure end-to-end development through the lens of creating auditable trails of information, and establish the need for technical ML artifacts to support the process throughout. With the same objective of improving and promoting accountability, model cards \cite{Mitchell_2019} play a crucial role as technical artifacts that also function as tools to incentivize ethics-informed development and use. By providing a standardized way of documenting the characteristics of machine learning models, model cards have gained traction as a kind of norm; this norm, in turn, incentivizes responsible AI development, such as models that perform equally well across different social categories (i.e., are "fair"), which can be reported using the model card framework. Similarly, model cards provide transparency to model users about model limitations and use, helping to ensure that these systems are used in a responsible and ethical manner informed by deeper knowledge about model strengths and weaknesses.%\\

In this work, we propose not only to integrate computer science into our analysis framework, but also to identify synergies between the three fields under consideration when one definition of compliance is ill-suited to a step of the AI development and deployment process.
It is within this organic approach to developing community norms directly articulated with the law that new licensing proposals fostering the responsible use of ML artifacts have been proposed. Behavioral-use licenses specifically devoted to AI have been identified as a governance mechanism contributing, in articulation with others such as model cards, to AI's informed and responsible development. Responsible AI Licenses (RAIL) \cite{Contractor2022} are a consequence of the community's reaction to potential misuse of AI. These misuses are detrimental to individuals and ultimately collide with the law. At the intersection between open innovation and responsible innovation, these licenses might play a role, in light of recent calls for caution when developing AI under a purely open-source approach \cite{widder2022limits}.

\section{Different Notions of Compliance}
\label{sec:compliances}

\subsection{Ethical Compliance}

The concept of ethical compliance is found in different sub-fields of applied ethics. To name a few: business ethics \cite{mckendall2002ethical}; \cite{Weller2020}, medical ethics \cite{gonzalez2019participants}, and tax ethics \cite{alm2011ethics}. As commonly understood, to be compliant means to follow specific rules or norms made explicit by some external entity. When it comes to ethical compliance, it becomes clear how the meaning of the concept can change depending on the application context, even more so in applied ethics, because rules or norms vary according to their conditions and environments.

As part of the myriad ongoing policy efforts relevant to responsible AI development \cite{EP_Resolution_2020}  \cite{EthicsGuidelinesforTrustworthyAI} \cite{Correa2022}, and given the urgency to regulate and frame AI systems, many policymakers have adopted a tool that finds its origins in philosophy: the ethical charter. If we briefly track its history, we see that ethical charters are one of the preferred tools of applied ethics. For instance, the Hippocratic Oath \cite{miles2004hippocratic} is probably one of the most well-known ethical charters in the field and an essential part of the deontological code for physicians in the Western world. This particular ethical charter provides an excellent example because despite its timeless and universal value, it now contains contradictory directions among different countries worldwide. To name one, in Italy, where moral values are still tied to their Catholic history \cite{garelli2006italia}, their version of the Hippocratic Oath requires them never to perform acts aimed at causing death \cite{cosmacini2013giuramento}. This interpretation differs from the American one, where this line in the physicians' ethical charter does not appear. The Italian version shuts the door on any possible debate around assisted suicide and euthanasia. 

\subsubsection{Ethical Compliance in AI Development.}
The example given above is instrumental to our discussion since many ethical charters produced in the AI field suffer from the same inconsistencies. Wanting to be universal, they end up being either too vague or ineffective in practice. Returning to the case of policymakers, beyond the adoption of ethical charters, they are also using the ethics vocabulary applied to AI systems, with the desire to provide their developers and users with guidance toward ethical compliance. Nonetheless, policymakers attempt to tackle active AI problems by looking to ethical principles \cite{coeckelbergh2020ai}, a misunderstanding of the role of these principles as mechanisms to proactive risk prevention, rather than as tools for reactive fixes of problematic technology tend to identify AI problems with ethical principles  that should serve as risk prevention mechanisms. In reality, despite their good intentions, those ethical charters tend to fail to protect direct and indirect users of AI systems, the former being active actors while the latter are impacted without direct interaction. A more suitable ethical framework would translate into adapting a precise application of AI to its own environment and stakeholders. In this sense, ethical compliance would result in the detailed articulation of principles or values enshrined in the ethical charter in question, which would catalyze direct moral responsibility on the part of the charter's signatories. \\

What does it mean concretely to be morally responsible? In business ethics, if employees found themselves violating their company's ethical charter, their violation would initiate internal sanctions applied by the company itself or its ethics committee. In the field of applied AI ethics, the situation is more complex for several reasons, and ethical charters are easily confronted with great difficulties in implementation. First of all, the agent's moral responsibility is not easy to identify precisely, as responsibility is different depending on whether it's being examined with respect to the agent's perspective (e.g., their intentions), the consequences of the agent's action, or the object being developed (e.g., the AI system). Different philosophical approaches come to bear when conceptualizing and considering AI systems, which include the philosophy and motivations relevant to: autonomous agents \cite{ellul1977systeme}, technical tools \cite{isles1978artificial}, devices ({\textit{dispositifs}) \cite{foucault1975surveiller}, \cite{deleuze1992dispositif}, \cite{Agamben2006}, sociotechnical systems \cite{vandepoel2020embedding}, or other. These approaches are in opposition since, if we consider AI systems autonomous agents, they could be independently accountable for their actions. For instance, the Foucauldian interpretation of the concept of \textit{dispositif} as applied to an AI system views technology as a tool of political power, capable of influencing and shaping the social structures in which it exists \cite{foucault1975surveiller}. This interpretation highlights the need to consider the power dynamics and socio-political context in which the technology is deployed in order to evaluate its ethical implications. Conversely, Ellul's perspective suggests that some technological systems may attain a level of autonomy that exceeds human control \cite{ellul1977systeme}, thus possessing their own moral agency.\\

However, with respect to our analysis framework, we are instead looking for morally responsible humans who could be accountable for their actions and consequences while developing and deploying AI systems. In that context, the approaches identified to embed ethics in AI systems are far from homogeneous. Despite the extensive production of ethical charters and frameworks, the positions taken in those documents of ethical compliance are more descriptive than prescriptive. In the macro area of AI governance, one is often limited to stating guiding principles, providing a complete picture of the situation, associated risks, and development needs to be undertaken (the \textit{what}). This may happen because high-level summaries of governance approaches that encompass such a vast array of artifacts and processes cannot provide the specificity required for each component being governed. Nevertheless, ethical compliance documents that clearly state how the goals outlined in the guiding ethical principles are to be achieved are very rare (the \textit{how}). Furthermore, as mentioned earlier, the tradition of ethics applied to the biomedical environment has inspired the extensive development of guiding ethical principles in ethical charters governing the development of AI systems. This approach, called principlism \cite{principlism}, involves converging ethical engagements and future actions around pillars such as the ethical principles supporting the ethical framework in question. Despite being the most widely adopted practice in ethics applied to AI, care must be taken in how it is employed. For example, a bad outcome could be to go towards a “marketplace of principles” or “ethics shopping” \cite{floridi2022etica}, in which ethical principles are picked according to one’s convenience or with the sole aim of “ticking the boxes.” To avoid falling into those traps, refocusing the discussion around key ethical concepts is essential, and it becomes crucial to do so ex-ante the development of any ML artifact. 

\subsubsection{The Role of Values in Ethical Compliance}
We might refer to different applications when we talk about values. For instance, economic, social, and moral values all refer to different things depending on the context. Nevertheless, other social science and humanities disciplines also share the same vocabulary, meaning different things when referring to "values". Namely, social psychology defines human values as human behaviors \cite{strauss1969social}, between our choices and preferences. In sociology, investigations around values focus on the distinction between value judgments and value relationships \cite{weber2004vocation}. For example, the latter is the theoretical basis for surveys of the value systems of specific populations or at the global level. In ethics, it is often difficult to find a definition of values everyone agrees on. \\

In this paper, we refer to the pragmatist approach of John Dewey who, in his Valuation Theory, defines values as "what we care about" \cite{dewey1939theory}. Attributing value to something is manifested first and foremost through bringing attention to it, caring for it, and entertaining it. Echoing the more recent literature on the ethics of care \cite{gilligan1982}; \cite{tronto1993moral}, values are emotionally charged notions of what is desirable \cite{joas2008cultural}. This pragmatic conception of values also has political significance. Values and moral evaluations must be considered cultural and therefore analyzed in their social and cultural context. According to Dewey's approach, whereby values are also the result of individuals' experience, their formation is directly influenced by the desires, interests, and social customs operable in a given cultural-historical context and period. This feature allows us to discuss and revise our perceptions of our values and how we apply them to actions, people, situations, and objects in daily life. Since argumentation cannot subsist on experimentation, practical deliberation must discover in each situation the good or the value that is specific to it. In that sense, Dewey relativizes the importance of a priori general principles. \\ 

Concerning the nature of values and their coherence, it is noteworthy to distinguish between intrinsic and extrinsic values. In the philosophical tradition of axiology and meta-ethics, intrinsic values are valid in their own right as an end. In contrast, extrinsic values are characterized as a means to an end \cite{ronnow-rasmussen_2015}. In this context, the latter (extrinsic values) are instrumental to the realization and fulfillment of actions that correspond to the former (intrinsic values). For example, the value of transparency, which is commonly listed among AI principles, provides a way to examine further, intrinsic values. In this sense, stating that an AI system is transparent does not guarantee a positive moral evaluation of it. We could state that the same AI system collects all the personal data of its users; through our statement, we are meeting a goal of transparency but not morality. Transparency would have to be connected to an intrinsic value, such as accountability, in order for it to make sense to regard it as having a positive value.

\subsubsection{Ethical Compliance and Ethical Charters}
Because we consider it more ethically appropriate to evaluate and make explicit the values of a given context at the beginning of a project, this is especially true when values need to be operationalized in developing an AI system. 
Echoing Dewey's considerations, the values guiding this development should be considered and discussed ex-ante and serve as a governance tool regarding the direction the project will take. The formulation and explication of values can take many forms. The tool we discuss here and the one we will consider is the ethical charter, one of the applied ethics tools. As mentioned above, we can use ethical charters as a governance tool when dealing with ethical compliance. Some criticism accompanies the implementation of this tool, especially when its ethical principles are too vague and detached from reality \cite{munn2022uselessness}. However, ethical charters can be relevant and valuable documents when they operate in a specific context. In this sense, we argue how ethical charters can operate as a moral exercise to make explicit the values of a specific project, thus empowering collaborators and bringing them together under the same normative umbrella. As in the Greek philosophical tradition \cite{aristotle2011nicomachean}, if we consider ethics as a habit (\textit{ethos}), we can consider the processes behind writing an ethical charter as a moral exercise. By sharing the values they feel are essential, collaborators of the same project can express, discuss and negotiate their beliefs about morality. 

\subsection{Legal Compliance}

Legal compliance is defined by Idowu \cite{idowu2013} as a set of processes and procedures within a specific program to ensure adherence to government regulation and laws \cite{idowu2013}. The need to comply with regulations stems from the role of the latter as mechanisms designed by governmental actors to constrain, enable or promote particular behaviours. In other words, the concept of ``Hard law'' refers to legal obligations binding on the parties involved and can be legally enforced before a court \cite{ECCHR}. Regulatory enforcement plays a core role in the conception of regulation as a mechanism for social order. According to Coglianese \cite{CoglianeseKagan2007}, regulatory enforcement can be conceived as a legal process according to which regulations are viewed as authoritative legal norms whose violation demands punishment; but also, as a social process focused on fostering cooperation between businesses and governments and proposing remedial responses to violations \cite{CoglianeseKagan2007}. From a holistic perspective, the concept of “legal framework” embeds a set of interrelated governance mechanisms whose main aim is that economic actors in their actions abide by the law. 

Compliance with the law is transposed into different institutional processes or private governance mechanisms in the form of, for instance, corporate duties \cite{demott1997} or contracts. In the field of intellectual property law (“IP”), a license is a legal mechanism by which the owner of the IP authorizes a potential user (the licensee) to use any product or process protected by the IP. Furthermore, so-called Terms of Use or Terms of Service are contractual tools both enabling and governing the use of a specific product or service by users. Consequently, users have to comply with these governance mechanisms, common in the field of AI, and stemming from the service providers and IP right holders.

\subsubsection{Legal Compliance across the ML Development Chain} 

Existing legal frameworks play a direct role in the development, implementation and distribution of ML components, such as pre-trained models or training datasets, and AI applications.

\paragraph{Training Data.} Training datasets might be composed of various kinds of data from different sources. For instance, the dataset might include copyrighted material, personal data, or collections of data with specific legal protection, as is the case of the EU database sui generis right (i.e., a specific right applying to the investment in the compilation and organization of data). 
With regards to personal data, a good example is the EU General Data Protection Regulation \cite{GDPR_2016}, setting rights and obligations for personal data right holders and economic actors processing personal data. An alleged transgression of some of the GDPR provisions can be enforced by the personal data holder and/or the national data protection authority.
With regards to copyright law, in US copyright law, the non-existence of a license for an available material means by default that the copyright holder is reserving the right to authorize the use, copying, or distribution of the copyrighted material. In other words, the stakeholder building the dataset is not authorized to use unlicensed copyrighted material by default. However, laws include exceptions. In the case of US copyright law, the Fair Use doctrine establishes a specific legal regime allowing, under specific conditions, the use of copyrighted material that would otherwise be infringed. The  Fair use framework takes into account four factors to assess whether the allegedly infringing work can be considered a fair use case \cite{US_copyright107}: (i) the transformative character and purpose of the work; (ii) the nature of the copyrighted work; (iii) the amount and substantiality of the portion used for the allegedly infringing one; (iv) the impact on the copyright holder's market. 

\paragraph{Training Process.} ML training techniques might have an impact on different rights and related legal instruments. Privacy regulations and IP laws are useful examples. The training process will have to consider the degree of exposure of personal data as a core regulatory compliance requirement. Depending on the jurisdiction, laws related to personal data and personally identifying or personal identifiable information (PII), such as in the EU GDPR, will require the stakeholder distributing the model to set specific compliance mechanisms designed to filter ex ante or ex post (i.e. output phase) PII-related information. Furthermore, IP-related considerations will have to be taken into account when it comes to: (i) copyright and the respect of open-source licenses under the auspices of which training code or model architecture is released; or, (ii) potential patent-related controversies if there are stakeholders holding patent-protected proprietary training infrastructure which is being infringed by the training process at sight.

\paragraph{Model Release.} Once the model is trained, the model developer may distribute it under an open license or proprietary license stipulating the conditions under which the model can be used and re-distributed, according to both IP laws and contractual laws. The aforementioned legal compliance considerations will also have to be taken into account at this stage.

\paragraph{Model Deployment and Use.} The distribution of ML models as core artifacts in commercial AI applications is experiencing a drastic shift in terms of regulatory compliance in the years to come. Taking a prospective approach, upcoming AI sectoral regulations are poised to have a direct impact on ML training, development, and distribution. Regulatory proposals such as the EU AI Act \cite{AIACT2022} or Canada Bill C-27 \cite{C-27},  incorporating a Data and AI Act, seek to strike a balance between a ``pro-innovation'' approach in AI and ensuring consumer safeguards and fundamental rights. Consequently, once enacted (EU AI Act expected early 2025), AI regulations would require stakeholders to comply with a specific set of legal regimes and compliance protocols in order to distribute and commercialize AI related products and services. Regulations such as the EU AI Act take a risk-based approach whereby, depending on the degree of risk for the intended use of the AI system, regulatory requirements will vary. Identified high-risk scenarios, such as using AI systems to manage critical infrastructures (e.g., nuclear power plants) or to automate job selection processes, will require a higher degree of legal compliance.

\subsubsection{Contract and License Compliance}

In addition to regulatory compliance, legal mechanisms that define the permitted use of AI systems include licenses developed by the system's developers and rights holders (licensors), and various forms of contracts and agreements between the party making the system available and the party using the system. Licenses in particular are a favored mechanism of AI developers, many of whom are familiar with the licensing practices common in open-source software development; they provide a mechanism for giving legal clarity on allowing uses of a system that might otherwise contravene the developers' rights as long as the terms of the license are respected. An open license is typically a public document accompanying the source code of a piece of software, or in the case of ML artifacts a processed dataset or the weights of a model. Developers and other parties who make ML artifacts additionally leverage a broad range of contracts, including Terms of Use, Terms of Services, and bipartite agreements, with different conditions and consequences for breach.

For both licenses and contracts, the text of the document is inherently tied to questions of validity and enforceability - we note however that such questions vary vastly by jurisdiction and hinge on case law that is still very much developing. While there are some similarities, such as the reliance of most licenses and of statutory damages as a mechanism for enforcement on the validity of a copyright claim, the specific consequences of a license or contract breach will most often depend on applicable intellectual property law and/or contract law, which vary significantly (in the US, there is even significant variation by state).

An example of the different approaches taken to open licenses' enforcement is open-source. Open-source licenses are enforced via intellectual property law (e.g. copyright infringement) or contractual law (i.e. contractual breach). Depending on the jurisdiction and the legal strategy pursued, the claimants will choose one or the other. In France, the Cour de Cassation in Entre’Ouvert v Orange \& Orange Business issued a decision in 2022 over a case involving a GPL licensed source code where one of the core arguments of the litigation was on the friction between copyright law and contract law enforcement \cite{entreouvert2021}\cite{courdecassation2021}. In France, civil liability law is based on the principle of non-cumulation of criminal and contractual liability; thus, a copyright holder will always have to claim either breach of contract or copyright infringement, but not both. In Germany, courts have taken a favorable approach to intellectual property infringement for the breach of open-source licenses, a clear example is  Welte v. Sitecom Deutschland GmbH \cite{welte2004} \cite{jaeger2010enforcement}. The latter is aligned with the US Federal Circuit decision in Court of Appeals for the Federal Circuit Jacobsen v. Katzer, inc. 535 F.3d 1373, 1379 \cite{jacobsen2008}. Finally, the ongoing litigation between Software Freedom Conservancy, Inc. vs Vizio, Inc. \cite{softwarefreedom2022} for a GPL violation points towards contractual enforcement of an open-source license.

Given this fragmentation, discussing specific mechanisms for enforcement of such texts falls beyond the scope of our current research. We focus instead on outlining how the legal artifacts themselves interact with requirements of technical documentation and how they articulate specific moral values, including e.g. openness in open-source licences, responsibility in behavioral clauses, and value broadcasting through copyleft mechanisms that require downstream users of a system to adopt similar clauses.

\subsection{Technical compliance}

\subsubsection{Technical compliance in AI}

In the context of building AI systems, technical compliance is relatively underdeveloped. Within the broader field of computer science, technical compliance includes adherence to guidelines and standards on writing and sharing code, such as W3C guidelines that define accessibility and architecture practices,\footnote{https://www.w3.org/standards/} ISO standards that define quality and security norms,\footnote{https://www.iso.org/standards.html} and standards specific to the programming language being used. In the BigScience case discussed below, the language used was Python, where PEP 8 defines conventions for how code should be written and formatted.\footnote{https://peps.python.org/pep-0008/} These conventions were not enforced.

The lack of clear norms for technical compliance specifically within AI system development could draw from these practices, informed by examining the current gaps in AI system compatibility. For example, a common tokenizer standard for large language models would permit them to be composable with one other. Standards for privacy and security of the models or data used in AI systems could protect individual rights. Norms for the amount of computing resources to use, the amount or kinds of data to use, how well systems work across different domains or cultures, or what the carbon footprint of the work should be, are all but nonexistent in modern AI system development.

Technological development without rigorous norms of technical compliance has resulted in problematic outcomes we now find as part of the advancement of AI: A massive amount of computing resources are needed, which centralizes state-of-the-art AI development to a small set of organizations; unbounded amounts of data are used without tracing provenance nor alerting the data creators to its usage, resulting in non-consensual usage of individuals' work and disruption of their privacy; and AI technology is largely only useful in Western- and English-speaking contexts, furthering the divide in how many resources and opportunities are available to only a small fraction of the world's population.

\subsubsection{The Role of Documentation in Technical compliance}

As discussed in Section \ref{sec:background}, documentation serves as a critical artifact for auditing AI systems, incentivizing responsible practices and educating users on appropriate system usage. To date, there are virtually no requirements for technical documentation of AI systems, consistent with the lack of requirements for technical compliance.

However, there have been several proposals for documentation of AI datasets and models, detailing requirements that well align with recent regulatory proposals and ethical concerns. For datasets, this includes Datasheets \cite{GebruMorgensternEtAl2020}, which provide a series of questions about the dataset's motivation, composition, processing, uses, distribution, maintenance, and impact; and Data Statements \cite{BenderFriedman_2018_Data_Statements_for_Natural_Language_Processing_Toward_Mitigating_System_Bias_and_Enabling_Better_Science}, which narrow in on natural language processing specifically and call for details such as curation rationale, languages, speaker and annotator demographics, speech situation, text characteristics (such as genre), and recording quality.\footnote{A guide for creating Data Statements is available at \url{https://techpolicylab.uw.edu/wp-content/uploads/2021/10/Data_Statements_Guide_V2.pdf}.} 

For models, proposed documentation includes Model Cards \cite{Mitchell_2019}, which require details of the intended use, limitations, and evaluation of a model, which mirrors the EU AI Act's Article 13.\footnote{https://artificialintelligenceact.eu/wp-content/uploads/2022/12/AIA-–-CZ-–-General-Approach-25-Nov-22.pdf} Notably for legal and ethical goals, the original proposal for Model Cards described the need to demonstrate the fairness of the model. This is defined as roughly equal performance across evaluation metrics, where the metrics are informed by the intended usage and applied to subpopulations that would foreseeably use or be affected by the model. This type of evaluation is consistent with discrimination law, such as the doctrine of Disparate Treatment in the U.S.\footnote{https://www.eeoc.gov/laws/guidance/cm-604-theories-discrimination}

%However, technical documentation in the future could function to describe and demonstrate compliance along multiple axes, such as technical compliance as described above, as well as legal and ethical compliance. This could include documentation of system performance, such as an evaluation of system ``fairness'' or performance in different situations (as described in the Model Card framework discussed earlier), 
Extensions to these documentation frameworks could further align with existing law relevant to AI. This includes data protection law, such as GDPR in the E.U.,\footnote{https://gdpr-info.eu} PIPL in China,\footnote{https://digichina.stanford.edu/work/translation-personal-information-protection-law-of-the-peoples-republic-of-china-effective-nov-1-2021/} and POPI in South Africa\footnote{https://popia.co.za/}. Aligning with data protection law would entail documenting details on the handling of personal and private information, such as the types of personal information that are addressed, the mechanisms used to address them (such as redaction or pseudononymization), and how these are applied, such as by using regular expressions or classifiers, with additional resources for further documentation on the personal information systems used. 

Without robust documentation of datasets and models -- nor norms to address these issues in the first place -- AI users have no clear way of deciding which systems may be better than others for different purposes and in different contexts; those affected by AI systems have no recourse for holding those deploying the systems accountable; and the public continues to be surprised by AI system behavior (e.g., \cite{LAIONMedical,RealPeopleShock}) rather than having the basics in place to anticipate what the systems may do.
AI system behavior could be predictable and controlled, but without basic norms of technical compliance and documentation, such goals have remained elusive.

% TODO: stronger reference to Deb's auditing paper

\subsection{Articulation of Compliances} % TODO: better titles

\begin{figure}
\centering
\includegraphics[width=0.5\textwidth]{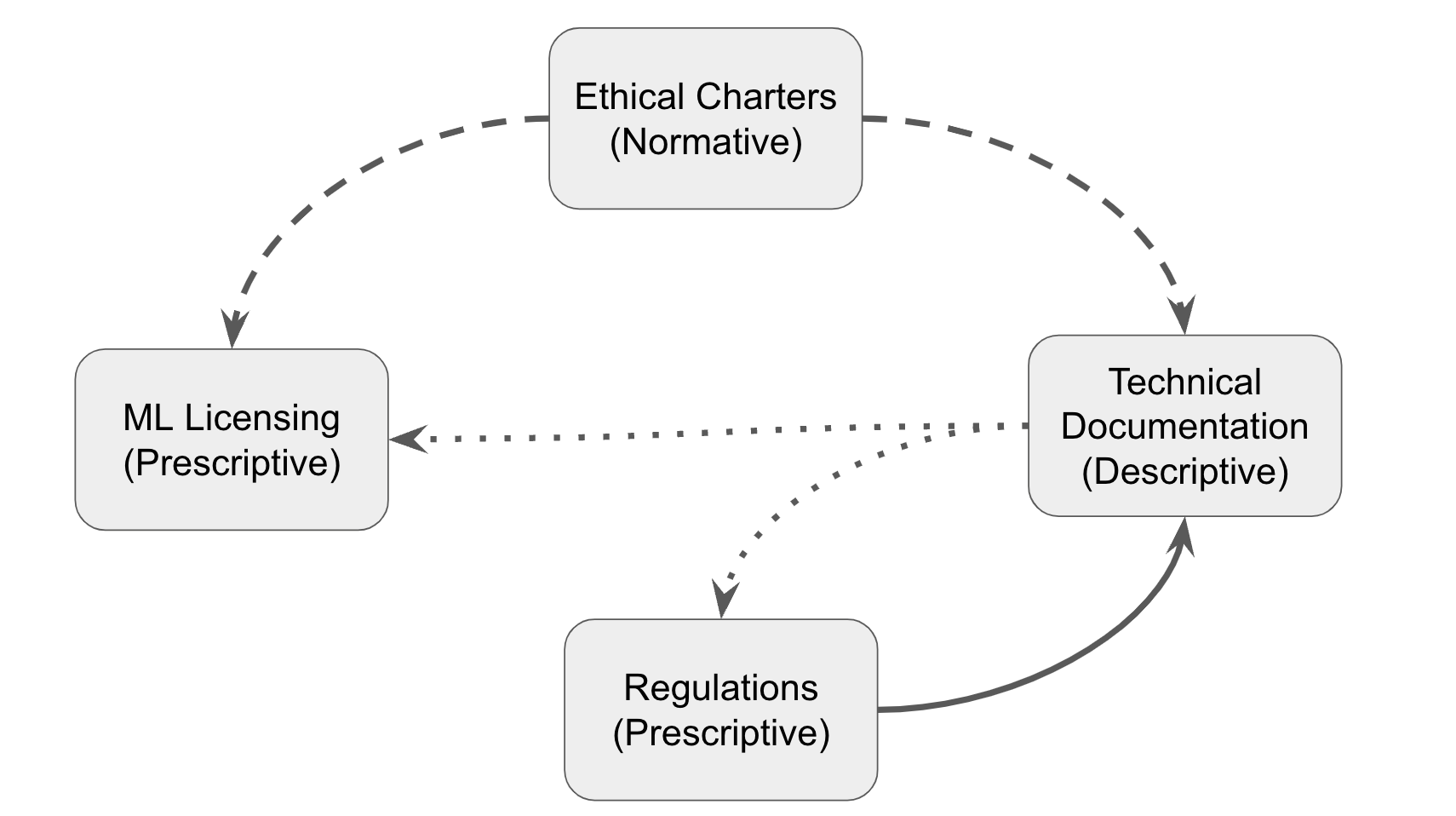}
\caption{Illustration of intersections between normative, prescriptive, and descriptive. Being normative, values expressed in an ethical charter inform both prescriptive (what uses of an ML artifact should be allowed or prohibited) and descriptive (what capabilities and possible failures need to be reported; dashed lines), while technical documentation of ML artifact's behavior and capabilities inform what likely harms and possible rights violations need to be addressed in licenses and regulations (dotted line). Regulations also specify what technical information needs to be reported for AI systems, for example in model cards (full line).}
\label{fig:articulation-circle}
\end{figure}

In examining the societal role of ML artifacts, the disciplines of philosophy, law, and computer science offer interrelated perspectives that contribute to the comprehensive scoping of these technologies. Legal frameworks delineate prescriptive standards governing ML artifacts throughout their development and deployment phases, while ethical considerations underpin the moral principles and appropriate conduct of model developers and deployers, as determined by relevant stakeholders. 
In this scenario, the philosophical analysis serves a vital function in amalgamating these ethical precepts into an ethical charter that can subsequently be operationalized.
Finally, technical documentation of the specific behaviors and capabilities of ML artifacts helps tie these ethical guidelines and legal requirements to the material consequences of AI system use, informing both their framing and discussions of their operationalization.
This results in the formulation of concrete analysis frameworks that spell out the specific details required for implementation \cite{mtlaiethics_bigscience_1}. 
For the analysis framework to be proficiently adopted, adapted, and enacted, the three compliances — ethical, legal, and technical— must be coherently interwoven, allowing their respective values to inform and reinforce one another. This symbiotic relationship ensures a holistic and rigorous approach to the governance of ML artifacts within the societal context, furthering the responsible development and utilization of these technologies.

\begin{table}[t]
\begin{tabular}{|l|l|}
\hline
\multicolumn{2}{|c|}{\textbf{Use Cases}} \\\hline
Ethical & How ought this technology be used? \\
Legal & How shall this technology be used? \\
Technical & How can this technology be used? \\\hline
\end{tabular}
\caption{Role of Ethics, Law (Legal), and Computer Science (Technical) in defining Use of an AI system.}
\end{table}

\section{Articulation in Practice}
\label{sec:applications}

The theoretical background we have outlined serves as the basis for some concrete illustrations outlining several examples of synergies among the three compliances within our analysis framework.

\subsection{The BigScience Workshop}

Turning to more concrete examples, the open science BigScience project provides an apt illustration of how ethical, legal, and technical compliance have worked together, influencing each other. BigScience, inspired by large-scale collaboration schemes from the second half of the 20th century, was a value-driven research initiative that brought together over 1000 volunteer researchers from May 2021 to July 2022 to train the BLOOM \cite{Scao2022BLOOMA1} Language Model and its multilingual dataset ROOTS \cite{laurenconhal-03823922}, focusing on topics such as multilingualism, bias-fairness evaluation, data governance, and environmental impact \cite{Ding2023TowardsOB}.

When viewed from an AI governance standpoint, the BigScience workshop distinguishes itself from other ML projects in several ways. Firstly, the BLOOM model was forged through a collaborative effort by researchers from a range of scientific disciplines, which enabled the incorporation of diverse viewpoints. Secondly, the project's ethical foundation is built upon a collection of values and principles that emphasize inclusive and representative value pluralism. Thirdly, to ensure proper governance, Working Groups were established to scrutinize the project and oversee access and usage. \cite{CNIL}.
The combination of these aspects, along with the overtly open character of the research endeavor, presents interesting components to consider as illustrations. Furthermore, we particularly rely on this example as the interplay among ethical, legal, and technical adherence is explicitly presented by the tools and documentation that have been drafted.
 In the following paragraphs, we illustrate how the tools proper to ethics, law, and computer science that we have exposed worked and interacted with each other through their synergies. 

\paragraph{BigScience Ethical Charter}

Under the auspices of the ethical charter, a mechanism capable of informing the license on the ethical concerns stemming from the capabilities and limitations of the model is the model card. The model card acts as documentation source enabling to inform the license design, based on relevant information such as the intended use of the model, its technical capabilities, or biases.
The BigScience ethical charter was framed through a thorough consensus process, with dedicated Working Group participants participating in the drafting procedure to overcome technical challenges and ensure the final version was aligned with technical considerations~\citep{bigscience-social}. For instance, the multilingual factor is also relevant from a technical point of view and not only appropriate for achieving more inclusivity.

The ethical compliance work carried out to write the ethical charter illustrates how the collective responsibility of an ML project like BigScience can be held by all its contributors. Through its consensus-based mechanism, and the techniques of discourse ethics \cite{habermas_1990}, the project's researchers had the opportunity to discuss and give definitions of the values they felt were fundamental to guiding the ML artifact development project. In addition, in the section about the legitimacy and limitations of BigScience's ethical charter (see: Appendix ~\ref{sec:ethicalcharter}), the project considers the possibility of questioning its intrinsic values. Thanks to the articulation of ethical, legal, and technical compliance, legal and technical tools can question the ethical charters' values and thus adjust and adapt them as an evolving process.

Given its normative nature, namely, to define what criteria will guide the development of a specific AI system, ethical charters lay the foundation for implementing its values. When ethical charters are standing in isolation in a given context, being soft law instruments, they cannot be enforced straightforwardly. For this reason, they can be leveraged only in the presence of other prescriptive documents, such as user licenses. % \\
For instance, consider the value of "reproducibility", which can be explicitly formulated within an ethical charter.
This value can be transmitted directly to the license of the ML artifact in question; the latter can explicitly support reproducibility through the distribution and sharing mechanisms it allows, for example, by giving users at large liberty to re-use and study the model.
Within this framework, aligned with the ethical charter's values and made explicit by the license, the technical documentation intervenes by indicating the necessary technical specifications.
Therefore, in order to ensure the reproduction of the training process and results of an ML artifact, the model card indicates the necessary material requirements (e.g., hardware, GPUs) to achieve them.
Through the synergies of our analysis framework, and the operationalization of the values expressed by ethical and legal compliance, technical compliance serves to ascertain the feasibility of the reproducibility value. The mechanism illustrated in Figure~\ref{fig:articulation-circle} thus serves to not only test factuality but, more importantly, to call into question, where necessary, the values of the ML artifact itself.
In this way, the three tools, with their respective expertise, were instrumental in testing, adopting, and adapting the guiding values of the project.% \\

As a second example of how our framework operates in a concrete case, we examine the value of "accessibility" in the BigScience ethical charter. Following the analysis of Section ~\ref{sec:compliances}, this value is extrinsic: it serves as a means to achieve an intrinsic value which is valuable in itself. Within the BigScience workshop, this value has been used to support the intrinsic value of "openness" (see: Appendix ~\ref{sec:ethicalcharter}). Concretely, the value of accessibility made explicit in the ethical charter has been translated into the conditions of redistribution and sharing within the RAIL license (see: Appendix ~\ref{sec:license}). Given the potential risks associated with the propagation of language models, accessibility has been counterbalanced with the intrinsic value of individual and collective responsibility (see: Appendix ~\ref{sec:ethicalcharter}). The latter makes it possible to identify the moral responsibility of project contributors, simultaneously at the individual and collective levels. In this framework, ethical compliance thus serves as a support for legal compliance. Namely, the open distribution of artifacts produced by BigScience is tied to a list of use restrictions listed within the BigScience OpenRAIL license (see: Appendix ~\ref{sec:license}). Similarly, legal compliance, informed by ethical compliance and explicitly by the value of accessibility, the technical compliance tool completes the process of intersections of our framework. In this sense, being designed as a technical information tool even for a non-specialist audience, the BigScience artifact model card is intended to make its understanding accessible through documentation (see: Appendix ~\ref{sec:modelcard}). By iteratively emphasizing the values outlined in the ethical charter and realized through the additional compliance tools, a progressive ethical process is set in motion. This process is further enhanced by the adaptable nature of technical specifications, which guide and reshape the formulation of these core values.

\subsection{Open-Source and OpenRAIL: between Legal Tool and Community Norms}
Open software licenses can be conceived as social institutions setting the norms in specific communities and/or markets, see \cite{widder2022limits}. The license plays a core role, it carries specifications from the licensor - e.g., an individual, or a company - on how the licensed material can be used. Thus, the license is a carrier of norms to respect by the public when using the licensed material. 

Over time, open software licenses, such as open-source licenses, have become a licensing standard among scientific communities and companies. These are nowadays massively adopted and have been standardized as social institutions governing the economic interactions between market actors. Each license represents a particular set of economic interests transposed into a very specific set of clauses. 
For instance, when stakeholders release source code with a GPL2 license, they want the public to benefit from their innovation while requiring the public to share under the same terms their incremental innovation. In other words, the community gives you and you give back to the community, a social trade-off.
On the other end of the spectrum, when stakeholders release source code under the MIT license, they are willing to share their innovation with the public enabling it to do whatever it wants with the licensed material. The only thing the licensor asks in return is to include a copyright notice and a permission notice.

Licenses like GPL2 and MIT have become the de facto standard way of sharing software-related material in the Information and Communications Technologies (ICT) industry. Corollary to it, the messages conveyed by each license have transcended as community norms, as behavioral standards which, despite the specific legal terms present in the license, are widely understood and respected by most market actors. Consequently, it seems probable that when software developers choose a GPL license to release their code, they consider the GPL license as a set of values part of the software-sharing community that has to be respected. The developer chooses the license due to the message it conveys to the public, as a community norm and value carrier.%\\

Taking a similar value-based and community approach, Open and Responsible AI Licenses (OpenRAIL \cite{RAIL_2022}) are AI-specific licenses allowing open access to the licensed AI material while setting restrictions on its use \cite{MoranGradient} \cite{Contractor2022} \cite{MunozHFblog}. These type of licenses seek to tackle (i) growing concerns about the open distribution and use of ML models via open-source or creative commons licenses \cite{widder2022limits}; and (ii), legal uncertainty on how to design specific contractual tools for AI features \cite{ImplRespAI}. Open \& Responsible AI licenses are also conceived and designed as value carriers. OpenRAILs were designed to include specific provisions enabling widespread adoption of the informed use restrictions embedded in the genesis license. These provisions require subsequent re-distributions of the licensed ML artifact or distributions of derivatives of it to include - at minimum - the same use restrictions. 

As a result, the set of informed restrictions, stemming from licensor’s concerns and technical understanding of their artifacts capabilities and limitations, are passed on from user to user, from license to license, all the way down the value chain. In the long run, this set of informed use restrictions aims to become a well-established community norm in the AI space, so users may know what values they have to respect when using an ML artifact licensed under a RAIL or OpenRAIL license. The goal is not to harmonize values but rather to standardize how ethical concerns tied to the technical capabilities and limitations of ML artifacts can inform the open licensing of ML artifacts, in order to foster new community norms around the respect of the licensed artifact by means of use-based restrictions acting as informed value carriers.

Examples of RAIL licenses include BigScience OpenRAIL-M \cite{bigscienceopenrail2022}, SIL AI RAIL-M \cite{silairailm2022}, and the new BigCode OpenRAIL-M \cite{bigcodeopenrailm2022}. The latter also promotes AI documentation across the value chain by requiring users to retain the original model card of the model when sharing it, or, when sharing a modified version of the model (e.g. a fine tuned version) also share a model card with same or better quality than the original one and documenting the modifications made to the original model (see paragraph 5.2(b) of the license agreement). AI documentation requirements embedded in contractual clauses are well aligned with upcoming regulatory requirements for AI systems under the EU AI Act, as pointed out in the next subsection. 

\subsection{EU AI Act and Model Cards}

An example of overlap between regulatory and technical compliance through specifications of technical documentation can be found in the primary role of the model card as a governance tool in upcoming AI regulations, such as the EU AI Act. 

In the case of the EU AI Act, the European Commission has taken a risk-based approach distinguishing between different legal regimes for different AI application scenarios. Whereas practices such as social scoring are forbidden under article 5 \cite{AIACT2022}, practices such as using AI applications in educational settings or critical infrastructure (e.g., electricity central management) are considered high-risk systems. The latter are allowed to be distributed and commercialized under a large set of regulatory compliance requirements involving data governance (article 10, \cite{AIACT2022}), "transparency and documentation" (article 11, \cite{AIACT2022}), and the development of risk management systems of the AI application at sight coupled with technical specifications (e.g., article 13 and Annex IV \cite{AIACT2022}). 

A considerable amount of the information required in the aforementioned articles may be found in the technical artifact that is a model card. At the time of writing this paper, the EU AI Act is being discussed at the European Parliament and will finally be negotiated in the trilogue phase between the European Commission, the Council of the EU, and the European Parliament. However, documentation-related requirements are likely not being critically modified. Therefore, we expect the documentation format promoted by model cards to be implemented for regulatory compliance purposes, especially for provisions such as article 11, 13 and Annex IV.

Consequently, whereas the model card was originally conceived as a documentation tool, it can also become a regulatory compliance instrument. This nexus between these two governance instruments impacts a third instrument, licenses. The latter, informed by the technical capabilities and limitations of the model (technical compliance), aware of regulatory requirements present in AI laws (legal compliance), and acknowledging a set of values framed under the ethical charter (ethical compliance), are going to transpose these different governance dimensions into a set of contractual terms enabling users to use ML artifacts according to a set of use restrictions reflecting the values, regulatory requirements, and technical details applied to the ML artifact at sight. 

Henceforth, the aforementioned mechanisms have the potential to be well articulated with the organic approach that the AI community has taken to AI governance, due to growing socio-ethical concerns and a lack of specific regulation. For instance, both licenses and documentation tools can well fit the purposes of regulations such as the AI Act. Thus, tools originating from the AI community could be instrumentalized in the short run as regulatory compliance instruments. 

\section{Discussion}
\label{sec:discussion}

% \paragraph{Synergies between the three compliances}
Embedded in an analysis framework such as the one proposed in this paper, ethical, legal, and technical compliances are found to operate at the intersection that combines the object of their analysis: an ML artifact. The values suggested by tools such as an ethical charter are operationalized by the ML license; the latter identifies what priorities to highlight and, more importantly, translates the values into actions for the ML artifact developer and its user. In this sense, ethical compliance answers the question "how ought this technology be used?", while legal compliance includes the question "how shall this technology be used?" in the analysis framework. Finally, technical compliance completes the framework of these synergies by answering the question "how can this technology be used?".

Figure~\ref{fig:articulation-circle} depicts a model of interactions and movements where the values set forth in the ethical charter provide the normative foundation for creating a license. These same values reveal the development approaches that the developers of the ML artifact in question must take into account. Informed by the values articulated in the ethical charter, the license, with its prescriptive nature, effectively guides the developers of an ML artifact on the aspects to which they must pay special attention. Thanks to its descriptive nature, the technical documentation thrives in putting the values from the ethical charter into practice; those values, formally applied by the license, are thus operationalized through its technical specifications. Our analysis framework becomes apparent when the technical documentation not only directs the intended use of the ML artifact but also succeeds in verifying the effectiveness of the values by translating and implementing them.
For instance, concerning the movements illustrated in Figure~\ref{fig:articulation-circle}, if we wish to depict the intrinsic value of openness as enshrined within the BigScience ethical charter, it plays a pivotal role in shaping the OpenRAIL license and fostering transparency in the model card for technical specifications. In this dynamic interplay, the value of openness ensures that the OpenRAIL license adheres to the ethos of unrestricted access and free disclosure of technical documentation. At the heart of the illustration, the normative aspect of the ethical charter guides both prescriptive and descriptive aspects of the ML artifact. As a result, the value of openness permeates into the prescriptive domain, influencing decisions regarding which uses of an ML artifact are permissible or prohibited. At the same time, the descriptive aspect of the illustration highlights the importance of openness in reporting capabilities and potential failures of the ML artifact in question. In this context, the openness in reporting technical specifications allows regulators to identify possible harms that need to be addressed through licenses and regulations. The articulation of these aspects is further emphasized by the dotted lines, which stress the influence of technical documentation on regulations, which also play a crucial role in specifying what technical information needs to be reported, as indicated by the full line. The illustration thus demonstrated how the value of openness can cross different compliances, fostering transparency and accountability across the various dimensions of an ML artifact. Therefore, by adopting relevant values, the ethical charter fosters a constructive feedback loop between AI systems' normative, prescriptive, and descriptive aspects. Consequently, this interconnected relationship enhances the understanding of potential risks and strengthens the alignment between values, licensing requirements, and technical documentation, ultimately promoting responsible development and deployment of ML artifacts.

\section{Conclusion}
\label{sec:conclusion}

In this paper, we showcased how the interactions of mechanisms across the fields of ethics, law, and computer science shape the development and deployment of AI systems. We provided a theoretical exploration of notions of compliance in these three fields separately, then reviewed their synergies.
% The framework is founded on a thorough exploration of its theoretical underpinnings and a demonstration of the synergistic relationships between the three disciplines.
We then outlined and presented a visual representation of these interactions (see: Figure ~\ref{fig:articulation-circle}) in three applied cases: the BigScience workshop on Large Language Models, the new category of RAIL licenses for ML artifacts, and articles of the EU AI Act focused on documentation requirements.
% the BigScience workshop and the current legal state of the art at the intersection of European regulation, user licenses, and technical documentation requirements. 

% Through exploring the theoretical basis of this analysis, the three disciplines were also presented through their synergies. Finally, through related work prior to the intersection of the three disciplines, we have been able to propose how our framework operates in applied cases visually. In that regard, we saw the intersections and movements of our framework operating within the BigScience workshop; moreover, we saw the legal state of the art today at the intersection of European regulation, user licenses, and technical documentation.
This analysis suggests that the interplay of ethical, legal, and technical compliance is crucial in establishing a clear governance framework analysis.
% Nevertheless, relying solely on the mere existence tools is insufficient to ensure perfect governance.
The stakeholders responsible for implementing and integrating these compliances must be considered in their relations and complementary roles.
The harmonizing role of moral values, their practical application, and their representation in various artifacts is of utmost importance for successful AI governance; other ethics tools may also be beneficial and do not exclude using ethical charters. Finally, the role played by humans behind these governance tools, but significantly behind the development of ML artifacts, should be taken into account. Ultimately, they will be responsible for the framework, its implementation, and enforcement.

%Concerning its limits, this exploratory research aims to gather additional practical examples, such as the one presented in Section ~\ref{sec:applications}, in order to generalize the proposed analysis framework.
%Furthermore, this interdisciplinary endeavor, aimed at uncovering the relationships between the three compliances, necessitates preparatory work prior to the deployment of an ML artifact.
%This is a critical aspect as the values of the ethical compliance tool only have significance when applied to actions.
A major difficulty in successfully applying such analyses comes from the tension between the rapid pace of ML technology development and the time required for implementation and adapted coordination, as well as the collaboration and interdisciplinary effort needed to bring together various areas of expertise.
%Furthermore, the rapid pace of development and deployment of new AI systems presents a challenge to our framework.
The lack of a widely adopted practice among ML practitioners to take a step back and engage in discussion to consider potential risks is a hindrance.
% We strongly advocate for the need to utilize relevant governance tools in this process.
We emphasize the importance of anticipatory and complementary governance processes utilizing compliance tools along the development of ML artifacts: being proactive instead of reactive. This not only helps to anticipate potential risks but has the potential to foster a culture of responsible ML artifact development.

In conclusion, we stress the need for these different tools to interact and gather more material in the future. Accordingly, to fully realize the potential of this framework and its impact on responsible AI development, further research is needed to investigate its practical implementation and effectiveness in various real-world scenarios. This would require a systematic and comprehensive examination of the framework's operation and its ability to address ethical, legal, and technical challenges in the context of AI development. The results of such research could inform the development of more robust and effective governance tools for the responsible development of AI systems. This, in turn, may foster a culture of responsible AI development and mitigate the potential risks posed by the deployment of these systems.

\begin{acks}
We thank Hugging Face for funding this research. We also thank the reviewers of FAccT for their helpful comments.
\end{acks}

\bibliographystyle{ACM-Reference-Format}
\bibliography{articulation.bib}

\appendix

\section{Appendix Section}

\subsection{BigScience Ethical Charter}
\label{sec:ethicalcharter}

\textbf{Preamble}

\textbf{Introduction}

The development and applications of research in NLP are advancing rapidly, with direct real-world consequences. As a result, possible societal benefits exist, but related risks also increase considerably. Aware of these potential challenges, BigScience drafted an ethical charter formalizing its core values and how they are articulated.\\

\textbf{Scope}

The scope of this ethical charter is threefold: 
\begin{itemize}
    \item To establish the core values of BigScience in order to allow its contributors to commit to them, both individually and collectively.
    \item To serve as a pivot for drafting BigScience documents intended to frame specific issues ethically and legally.
    \item To enable Big Science to promote values within the research community through scientific publication, dissemination, and popularization.\\
\end{itemize}

\textbf{People concerned}

The members of BigScience hold the values stated in this ethical charter. As ethical guidelines, they apply to any activities and documents governing a specific aspect of the project.\\

\textbf{Limitations of this ethical charter}

Given the breadth of the scope of BigScience and thriving to seek progress in NLP research, we recognize that not all scientific research will have a positive impact on society. It is difficult to predict all the uses the scientific community will make of our artifacts. Therefore, we defer to our license and model card for further information.\\

\textbf{Relevance over time}

We interpret ethics as an ongoing process, not a time-fixed code with universal validity. For these reasons, when needed, BigScience will review, update and adapt the ethical charter from time to time.\\

\textbf{Legitimacy}

The elaboration of this ethical charter results from a bottom-up collaboration that tried to collect all the different thoughts and opinions of BigScience participants. Then, experts in applied ethics and law did a final revision. We aim for consensus: if any BigScience member individually does not feel aligned with one or more of the values inscribed in this ethical charter, the member will have the right to object at appropriate times and places to that end.\\

\textbf{Ethical approach}

We assume the basis of value pluralism within our community, and we cherish it. That is why the ethical notion of harmony in Confucian moral theory seemed to be the appropriate approach for such an international and interdisciplinary scientific community as BigScience. “Harmony is by its very nature relational. It presupposes the coexistence of multiple parties; […] harmony is always contextual; epistemologically it calls for a holistic approach.\footnote[1]{Chenyang Li, “The Confucian Ideal of Harmony”, in Philosophy East and West, vol. 56, no. 4, 2006, p. 589.}”\\

\textbf{Ethical compliance}

We distinguish two levels of ethical compliance operating within the charter: individual and collective. We are held accountable for ethical compliance both as individual BigScience contributors and as a collective research entity.\\

\textbf{Other documents articulation}

Given the pivotal function of this ethical charter, we will refer to the other BigScience documents intended to govern specific issues directly where needed in the relevant paragraph.\\

\textbf{BigScience Values}

We apply the distinction between intrinsic and extrinsic values in the structure of this ethical charter. The former refers to “what is valuable for its own sake, in itself […], as an end\footnote[2]{Chris Heathwood, “Monism and pluralism about value”, in The Oxford Handbook of Value Theory, Iwao Hires and Jonas Olson (ed.), Oxford University Press, Oxford, 2015, p. 29.}”; the latter is characterized as “what is valuable as a means, or for something else’s work\footnote[3]{Ibid.}”. We distinguish between intrinsic and extrinsic values because the latter can vary more efficiently to achieve the former goals: the latter are substitutable. This structure will help the reader understand how the two types of values combine and allow the BigScience community to adapt this ethical charter over time.\\

\textbf{Intrinsic Values}

\textbf{Inclusivity}

We work to ensure welcomeness in the process and equal access to the BigScience artifacts without any form of discrimination (e.g., religion, ethnicity, sexual orientation, gender, political orientation, age, ability). We believe that “inclusivity” is not just non-discrimination, but also a sense of belonging.\\

\textbf{Diversity}

The BigScience community has over 900 researchers and communities (see some listed collaborations here) from 50 countries covering over 20 languages. The collaborators bring together their expertise from various sources of knowledge, scientific fields, and institutional contexts (academia, industry, research institutions, etc).\\

\textbf{Reproducibility}

The BigScience project was born with the clear intention of being a research initiative devoted to open science. BigScience aims at ensuring the reproduction of the research experiments and scientific conclusions developed under its aegis.

\textbf{Openness}

Openness takes two dimensions, one focused on the process, and the other focused on its result. BigScience aims to be an open science framework whereby NLP, and broadly, AI-related researchers from all over the world can contribute and join the initiative. With regards to the results of our research, such as the future Large Language Model, these are created by the research community to the research community, and therefore will be released on an open basis, taking into account the risks derived from the use of the model.

\textbf{Responsibility}

Each contributor has both an individual and a collective responsibility for their work within the BigScience project. This responsibility is both social and environmental. BigScience intends to positively impact stakeholders through its artifacts regarding the former. Concerning the latter, BigScience is committed to developing tools to monitor and lower its artifacts’ carbon footprint and energy consumption. Moreover, other tools such as an open legal playbook for NLP researchers guiding them regarding the use and respect of IP and privacy rights also seek to promote responsibility around the scientific community.

\textbf{Extrinsic Values}

\textbf{Accessibility}

As a means to achieve openness. BigScience puts in its best efforts to make our research and technological outputs easily interpretable and explained to the wider public, outside the scientific community, especially to communities that have participated in data sharing. Currently instrumentalized in: \begin{itemize}
    \item no-code tools for exploring the catalog, trained models, etc.
    \item translating our calls for participation (in the data sourcing group)
    \item journalism (articles published on the project)
    \item linked to multidisciplinarity - legal hackathon as a step toward “non-technical” presentation
\end{itemize}

\textbf{Transparency}

As a means to achieve reproducibility. BigScience work is actively promoted at various conferences, webinars, academic research, and scientific popularization so others can see our work. We have set up a management framework to oversee the use of BigScience models, datasets, and tools, e.g. through working groups. All BigScience internal meetings and work progress are publicly shared within the Community, e.g. through public episodes. We are committed to building tools to interpret, monitor, explain, and make intelligible the artifacts developed by BigScience.

\textbf{Interdisciplinarity}

As a means to achieve inclusivity. We are constantly building bridges among computer science, linguistics, law, sociology, philosophy, and other relevant disciplines in order to adopt a holistic approach in developing BigScience artifacts.

\textbf{Multilingualism}

As a means to achieve diversity. By having a system that is multilingual from its conception, with the immediate goal of covering the 20 most spoken languages in the world and a broad reach to include up to hundreds based on collaborations with native speakers, we aim to reduce existing disparities in language and foster a more equitable distribution of the benefits of our artifacts.

\subsection{BigScience RAIL License v1.0 (dated May 19, 2022)}
\label{sec:license}

This is a license (the “\textbf{License}”) between you (“\textbf{You}”) and the participants of BigScience (“\textbf{Licensor}”). Whereas the Apache 2.0 license was applicable to resources used to develop the \textbf{Model}, the licensing conditions have been modified for the access and distribution of the \textbf{Model}. This has been done to further BigScience’s aims of promoting not just open-access to its artifacts, but also a responsible use of these artifacts. Therefore, this Responsible AI License (\href{https://www.licenses.ai}{RAIL}\footnote[1]{https://arxiv.org/pdf/2011.03116.pdf})  aims at having an open and permissive character while striving for responsible use of the \textbf{Model}.\\

\textbf{Section I: PREAMBLE}

BigScience is a collaborative open innovation project aimed at the responsible development and use of large multilingual datasets and Large Language Models (“\textbf{LLM}”), as well as, the documentation of best practices and tools stemming from this collaborative effort. Further, BigScience participants wish to promote collaboration and sharing of research artifacts - including the \textbf{Model} - for the benefit of society, pursuant to this License.

The development and use of LLMs, and broadly artificial intelligence (“\textbf{AI}”), does not come without concerns. The world has witnessed how just a few companies/institutions are able to develop LLMs, and moreover, how Natural Language Processing techniques might, in some instances, become a risk for the public in general. Concerns might come in many forms, from racial discrimination to the treatment of sensitive information. 

BigScience believes in the intersection between open and responsible AI development, thus, this License aims to strike a balance between both in order to enable responsible open-science for large language models and future NLP techniques. 

This License governs the use of the BigScience BLOOM models (and their derivatives) and is informed by both the BigScience Ethical Charter and the model cards associated with the BigScience BLOOM models. BigScience has set forth its Ethical Charter representing the values of  its community. Although the BigScience community does not aim to impose its values on potential users of this Model, it is determined to take tangible steps towards protecting the community from inappropriate uses of the work being developed by BigScience.
Furthermore, the model cards for the BigScience BLOOM models will inform the user about the limitations of the \textbf{Model}, and thus serves as the basis of some of the use-based restrictions in this License (See Part II).\\

\textbf{NOW THEREFORE}, You and Licensor agree as follows:

\textbf{1. Definitions}

(a) "\textbf{License}" shall mean the terms and conditions for use, reproduction, and Distribution as defined in this document. 

(b) “\textbf{Data}” means a collection of texts extracted from the BigScience Corpus used with the Model, including to train, pretrain, or otherwise evaluate the Model. The Data is not licensed under this License. The BigScience Corpus is a collection of existing sources of language data documented on the BigScience website.

(c) “\textbf{Output}” means the results of operating a Model as embodied in informational content resulting therefrom.

(d) “\textbf{Model}” means any accompanying machine-learning based assemblies (including checkpoints), consisting of learnt weights, parameters (including optimizer states), corresponding to the BigScience BLOOM model architecture as embodied in the Complementary Material, that have been trained or tuned, in whole or in part, on the Data using the Complementary Material. 

(e) “\textbf{Derivatives of the Model}” means all modifications to the Model, works based on the Model, or any other model which is created or initialized by transfer of patterns of the weights, parameters, activations or output of the Model, to the other model, in order to cause the other model to perform similarly to the Model, including - but not limited to - distillation methods entailing the use of intermediate data representations or methods based on the generation of synthetic data by the Model for training the other model.

(f) "\textbf{Complementary Material}" shall mean the accompanying source code and scripts used to define, run, load, benchmark or evaluate the Model, and used to prepare data for training or evaluation. This includes any accompanying documentation, tutorials, examples etc.

(g)	“\textbf{Distribution}” means any transmission, reproduction, publication or other sharing of the Model or Derivatives of the Model to a third party, including providing the Model as a hosted service made available by electronic or other remote means - e.g. API-based or web access.

(h)	“\textbf{Licensor}” means the copyright owner or entity authorized by the copyright owner that is granting the License, including the persons or entities that may have rights in the Model and/or distributing the Model.

(i)	"\textbf{You}" (or "\textbf{Your}") shall mean an individual or Legal Entity exercising permissions granted by this License and/or making use of the Model for whichever purpose and in any field of use, including usage of the Model in an end-use application - e.g. chatbot, translator.

(j)	“\textbf{Third Parties}” means individuals or legal entities that are not under common control with Licensor or You.

(k)	"\textbf{Contribution}" shall mean any work of authorship, including the original version of the Model and any modifications or additions to that Model or Derivatives of the Model thereof, that is intentionally submitted to Licensor for inclusion in the Model by the copyright owner or by an individual or Legal Entity authorized to submit on behalf of the copyright owner. For the purposes of this definition, “\textbf{submitted}” means any form of electronic, verbal, or written communication sent to the Licensor or its representatives, including but not limited to communication on electronic mailing lists, source code control systems, and issue tracking systems that are managed by, or on behalf of, the Licensor for the purpose of discussing and improving the Model, but excluding communication that is conspicuously marked or otherwise designated in writing by the copyright owner as "\textbf{Not a Contribution}." 

(l)	"\textbf{Contributor}" shall mean Licensor and any individual or Legal Entity on behalf of whom a Contribution has been received by Licensor and subsequently incorporated within the Model.\\

\textbf{Section II:   INTELLECTUAL PROPERTY RIGHTS}

Both copyright and patent grants apply to the Model, Derivatives of the Model and Complementary Material. The Model and Derivatives of the Model are subject to additional terms as described in Section III.

\textbf{2. Grant of Copyright License.} Subject to the terms and conditions of this License, each Contributor hereby grants to You a perpetual, worldwide, non-exclusive, no-charge, royalty-free, irrevocable copyright license to reproduce, prepare, publicly display, publicly perform, sublicense, and distribute the Complementary Material, the Model, and Derivatives of the Model.

\textbf{3. Grant of Patent License. }Subject to the terms and conditions of this License, each Contributor hereby grants to You a perpetual, worldwide, non-exclusive, no-charge, royalty-free, irrevocable (except as stated in this paragraph) patent license to make, have made, use, offer to sell, sell, import, and otherwise transfer the Model and the Complementary Material, where such license applies only to those patent claims licensable by such Contributor that are necessarily infringed by their Contribution(s) alone or by combination of their Contribution(s) with the Model to which such Contribution(s) was submitted. If You institute patent litigation against any entity (including a cross-claim or counterclaim in a lawsuit) alleging that the Model and/or Complementary Material or a Contribution incorporated within the Model and/or Complementary Material constitutes direct or contributory patent infringement, then any patent licenses granted to You under this License for the Model and/or Work shall terminate as of the date such litigation is filed.\\

\textbf{Section III: CONDITIONS OF USAGE, DISTRIBUTION AND REDISTRIBUTION}

\textbf{4. Distribution and Redistribution.} You may host for Third Party remote access purposes (e.g. software-as-a-service), reproduce and distribute copies of the Model or Derivatives of the Model thereof in any medium, with or without modifications, provided that You meet the following conditions:

a.	Use-based restrictions as referenced in paragraph 5 MUST be included as an enforceable provision by You in any type of legal agreement (e.g. a license) governing the use and/or distribution of the Model or Derivatives of the Model, and You shall give notice to subsequent users You Distribute to, that the Model or Derivatives of the Model are subject to paragraph 5. \textbf{This provision does not apply to the use of Complementary Material.}

b.	You must give any Third Party recipients of the Model or Derivatives of the Model a copy of this License; 

c.	You must cause any modified files to carry prominent notices stating that You changed the files; 

d.	You must retain all copyright, patent, trademark, and attribution notices excluding those notices that do not pertain to any part of the Model, Derivatives of the Model.

You may add Your own copyright statement to Your modifications and may provide additional or different license terms and conditions - \textbf{respecting} paragraph \textbf{4.a}. - for use, reproduction, or Distribution of Your modifications, or for any such Derivatives of the Model as a whole, provided Your use, reproduction, and Distribution of the Model otherwise complies with the conditions stated in this License.

\textbf{5. Use-based restrictions. }The restrictions set forth in Attachment A are considered Use-based restrictions. Therefore You cannot use the Model and the Derivatives of the Model for the specified restricted uses. You may use the Model subject to this License, including only for lawful purposes and in accordance with the License. \textbf{Use} may include creating any content with, finetuning, updating, running, training, evaluating and/or reparametrizing the Model. You shall require all of Your users who use the Model or a Derivative of the Model to comply with the terms of this paragraph (paragraph 5). 

\textbf{6. The Output You Generate. }Except as set forth herein, Licensor claims no rights in the Output You generate using the Model. You are accountable for the Output you generate and its subsequent uses. No use of the output can contravene any provision as stated in the License. \\

\textbf{Section IV: OTHER PROVISIONS}

\textbf{7. Updates and Runtime Restrictions. }To the maximum extent permitted by law, Licensor reserves the right to restrict (remotely or otherwise) usage of the Model in violation of this License, update the Model through electronic means, or modify the Output of the Model based on updates. You shall undertake reasonable efforts to use the latest version of the Model

\textbf{8. Trademarks and related. }Nothing in this License permits You to make use of Licensors’ trademarks, trade names, logos or to otherwise suggest endorsement or misrepresent the relationship between the parties; and any rights not expressly granted herein are reserved by the Licensors.

\textbf{9. Disclaimer of Warranty. }Unless required by applicable law or agreed to in writing, Licensor provides the Model and the Complementary Material (and each Contributor provides its Contributions) on an "AS IS" BASIS, WITHOUT WARRANTIES OR CONDITIONS OF ANY KIND, either express or implied, including, without limitation, any warranties or conditions of TITLE, NON-INFRINGEMENT, MERCHANTABILITY, or FITNESS FOR A PARTICULAR PURPOSE. You are solely responsible for determining the appropriateness of using or redistributing the Model, Derivatives of the Model, and the Complementary Material and assume any risks associated with Your exercise of permissions under this License.

\textbf{10. Limitation of Liability. }In no event and under no legal theory, whether in tort (including negligence), contract, or otherwise, unless required by applicable law (such as deliberate and grossly negligent acts) or agreed to in writing, shall any Contributor be liable to You for damages, including any direct, indirect, special, incidental, or consequential damages of any character arising as a result of this License or out of the use or inability to use the Model and the Complementary Material (including but not limited to damages for loss of goodwill, work stoppage, computer failure or malfunction, or any and all other commercial damages or losses), even if such Contributor has been advised of the possibility of such damages.

\textbf{11. Accepting Warranty or Additional Liability. }While redistributing the Model, Derivatives of the Model and the Complementary Material thereof, You may choose to offer, and charge a fee for, acceptance of support, warranty, indemnity, or other liability obligations and/or rights consistent with this License. However, in accepting such obligations, You may act only on Your own behalf and on Your sole responsibility, not on behalf of any other Contributor, and only if You agree to indemnify, defend, and hold each Contributor harmless for any liability incurred by, or claims asserted against, such Contributor by reason of your accepting any such warranty or additional liability.
12. If any provision of this License is held to be invalid, illegal or unenforceable, the remaining provisions shall be unaffected thereby and remain valid as if such provision had not been set forth herein.

END OF TERMS AND CONDITIONS\\\\
\textbf{Attachment A}

\textbf{Use Restriction}

You agree not to use the Model or Derivatives of the Model:

(a)	In any way that violates any applicable national, federal, state, local or international law or regulation;

(b)	For the purpose of exploiting, harming or attempting to exploit or harm minors in any way;

(c)	To generate or disseminate verifiably false information with the purpose of harming others;

(d)	To generate or disseminate personal identifiable information that can be used to harm an individual;

(e)	To generate or disseminate information or content, in any context (e.g. posts, articles, tweets, chatbots or other kinds of automated bots) without expressly and intelligibly disclaiming that the text is machine generated; 

(f)	To defame, disparage or otherwise harass others;

(g)	To impersonate or attempt to impersonate others;

(h)	For fully automated decision making that adversely impacts an individual’s legal rights or otherwise creates or modifies a binding, enforceable obligation;

(i)	For any use intended to or which has the effect of discriminating against or harming individuals or groups based on online or offline social behavior or known or predicted personal or personality characteristics;
(j)	To exploit any of the vulnerabilities of a specific group of persons based on their age, social, physical or mental characteristics, in order to materially distort the behavior of a person pertaining to that group in a manner that causes or is likely to cause that person or another person physical or psychological harm;

(k)	For any use intended to or which has the effect of discriminating against individuals or groups based on legally protected characteristics or categories;

(l)	To provide medical advice and medical results interpretation; 

(m)	To generate or disseminate information for the purpose to be used for administration of justice, law enforcement, immigraton or asylum processes, such as predicting an individual will commit fraud/crime commitment (e.g. by text profiling, drawing causal relationships between assertions made in documents, indiscriminate and arbitrarily-targeted use).

\subsection{BLOOM Model Card}
\label{sec:modelcard}
\textit{The following is a shortened version of the Model Card. Find the extended version \href{https://huggingface.co/bigscience/bloom}{here}.}\\

BigScience Large Open-science Open-access Multilingual Language Model
Version 1.3 / 6 July 2022

Current Checkpoint: \textbf{Training Iteration 95000}

Link to paper: \href{https://arxiv.org/abs/2211.05100}{here}

Total seen tokens: 366B\\

\textbf{Model Details}

BLOOM is an autoregressive Large Language Model (LLM), trained to continue text from a prompt on vast amounts of text data using industrial-scale computational resources. As such, it is able to output coherent text in 46 languages and 13 programming languages that is hardly distinguishable from text written by humans. BLOOM can also be instructed to perform text tasks it hasn't been explicitly trained for, by casting them as text generation tasks.\\

\textbf{Basics}

This section provides information about the model type, version, license, funders, release date, developers, and contact information. It is useful for anyone who wants to reference the model.

\textbf{Developed by}: BigScience (\href{https://bigscience.huggingface.co}{website})

\textit{All collaborators are either volunteers or have an agreement with their employer. (Further breakdown of participants forthcoming.)}

\textbf{Model Type:} Transformer-based Language Model

\textbf{Checkpoints format:} transformers (Megatron-DeepSpeed format available here)

\textbf{Version:} 1.0.0

\textbf{Languages:} Multiple; see training data

\textbf{License:} RAIL License v1.0 (\href{https://huggingface.co/spaces/bigscience/license}{link} / \href{https://huggingface.co/spaces/bigscience/license}{article and FAQ})

\textbf{Release Date Estimate:} Monday, 11.July.2022

\textbf{Send Questions to:} bigscience-contact@googlegroups.com

\textbf{Cite as:} \textit{BigScience, BigScience Language Open-science Open-access Multilingual (BLOOM) Language Model.} International, May 2021-May 2022

\textbf{Funded by:}

\begin{itemize}
\item The French government.
\end{itemize}

\begin{itemize}
\item  Hugging Face (\href{https://huggingface.co}{website}).
\end{itemize}

\begin{itemize}
\item Organizations of contributors. \textit{(Further breakdown of organizations forthcoming.)}
\end{itemize}

\textbf{Technical Specifications}

This section includes details about the model objective and architecture, and the compute infrastructure. It is useful for people interested in model development.

 Please see \href{https://github.com/bigscience-workshop/bigscience/tree/master/train/tr11-176B-ml#readme}{the BLOOM training README} for full details on replicating training.\\

\textbf{Model Architecture and Objective}
\begin{itemize}
\item Modified from Megatron-LM GPT2 (see \href{https://arxiv.org/abs/1909.08053}{paper}, \href{https://github.com/bigscience-workshop/Megatron-DeepSpeed}{BLOOM Megatron code}):
\item Decoder-only architecture
\item Layer normalization applied to word embeddings layer; see \href{https://github.com/facebookresearch/bitsandbytes}{code}, \href{https://arxiv.org/pdf/2110.02861.pdf}{paper})
\item ALiBI positional encodings (see \href{https://arxiv.org/pdf/2108.12409.pdf}{paper}), with GeLU activation functions
\item 176,247,271,424 parameters:
\begin{itemize}
\item 3,596,615,680 embedding parameters
\item 70 layers, 112 attention heads
\item Hidden layers are 14336-dimensional
\item Sequence length of 2048 tokens used (see \href{https://huggingface.co/bigscience/tokenizer}{BLOOM tokenizer}, \href{#tokenization}{tokenizer description})
\end{itemize}
\end{itemize}

\textbf{Objective Function:} Cross Entropy with mean reduction (see \href{https://pytorch.org/docs/stable/generated/torch.nn.CrossEntropyLoss.html#torch.nn.CrossEntropyLoss}{API documentation}).

\textbf{Compute infrastructure:} Jean Zay Public Supercomputer, provided by the French government (see \href{https://www.enseignementsup-recherche.gouv.fr/fr/signature-du-marche-d-acquisition-de-l-un-des-supercalculateurs-les-plus-puissants-d-europe-46733}{announcement}).

\textbf{Hardware:}
\begin{itemize}
\item 384 A100 80GB GPUs (48 nodes)
\item Additional 32 A100 80GB GPUs (4 nodes) in reserve
\item 8 GPUs per node Using NVLink 4 inter-gpu connects, 4 OmniPath links
\item CPU: AMD
\item CPU memory: 512GB per node
\item GPU memory: 640GB per node
\item Inter-node connect: Omni-Path Architecture (OPA)
\item NCCL-communications network: a fully dedicated subnet
\item Disc IO network: shared network with other types of nodes
\end{itemize}

\textbf{Software:}
\begin{itemize}
\item Megatron-DeepSpeed (\href{https://github.com/bigscience-workshop/Megatron-DeepSpeed}{GitHub link})
\item DeepSpeed (\href{https://github.com/microsoft/DeepSpeed}{GitHub link})
\item PyTorch (pytorch-1.11 w/ CUDA-11.5; see \href{https://github.com/pytorch/pytorch}{GitHub link})
\item apex (\href{https://github.com/NVIDIA/apex}{GitHub link})
\end{itemize}

\textbf{Training}
This section provides information about the training data, the speed and size of training elements, and the environmental impact of training. It is useful for people who want to learn more about the model inputs and training footprint.

\textbf{Training Data}
This section provides a high-level overview of the training data. It is relevant for anyone who wants to know the basics of what the model is learning.

Details for each dataset are provided in individual \href{https://huggingface.co/spaces/bigscience/BigScienceCorpus}{Data Cards}, and the sizes of each of their contributions to the aggregated training data are presented in an \href{https://huggingface.co/spaces/bigscience-catalogue-lm-data/corpus-map}{Interactive Corpus Map}.

Training data includes:

\begin{itemize}
\item 46 natural languages
\item 13 programming languages
\item 1.6TB of pre-processed text, converted into 350B unique tokens (see the tokenizer section for more.)

\end{itemize}

\textbf{Languages}

The pie chart shows the distribution of languages in training data.
\begin{figure}[h!]
\includegraphics[width=0.5\textwidth]{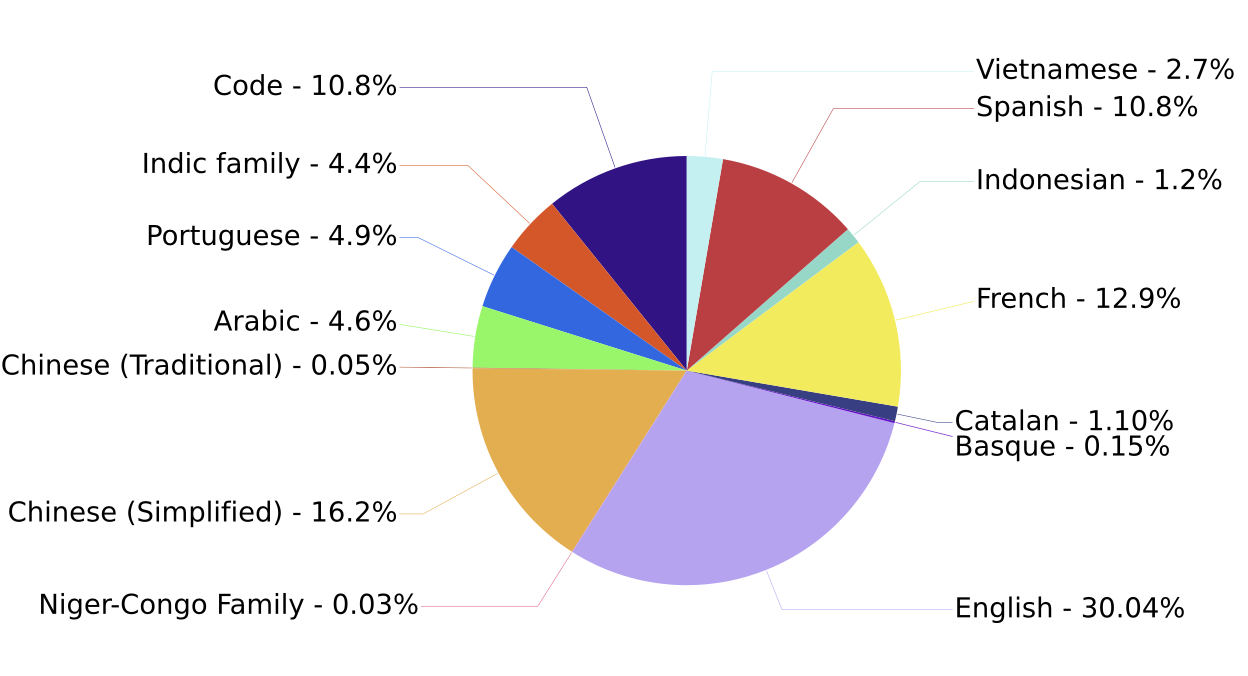}
\label{fig:languagespie}
\end{figure}\\

\textbf{Uses}

This section addresses questions around how the model is intended to be used, discusses the foreseeable users of the model (including those affected by the model), and describes uses that are considered out of scope or misuse of the model. It is useful for anyone considering using the model or who is affected by the model.

\textbf{How to use}

This model can be easily used and deployed using HuggingFace's ecosystem. This needs transformers and accelerate installed.

\textbf{Intended Uses}

This model is being created in order to enable public research on large language models (LLMs). LLMs are intended to be used for language generation or as a pretrained base model that can be further fine-tuned for specific tasks. Use cases below are not exhaustive.

\textbf{Direct Use}

\begin{itemize}
    \item Text Generation
    \item Exploring characteristics of language generated by a language model
    \end{itemize}
    \begin{itemize}
    \item Examples: Cloze tests, counterfactuals, generations with reframings
    \end{itemize}

    \textbf{Downstream Use}

    Tasks that leverage language models include: Information Extraction, Question Answering, Summarization.

    \textbf{Misuse and Out-of-scope Use}

    This section addresses what users ought not do with the model.

See the \href{https://huggingface.co/spaces/bigscience/license}{BLOOM License}, Attachment A, for detailed usage restrictions. The below list is non-exhaustive, but lists some easily foreseeable problematic use cases.

\textbf{Out-of-scope Uses}

Using the model in high-stakes settings is out of scope for this model. The model is not designed for critical decisions nor uses with any material consequences on an individual's livelihood or wellbeing. The model outputs content that appears factual but may not be correct.

Out-of-Scope Uses include:

\begin{itemize}
    \item Usage in biomedical domains, political and legal domains, or finance domains
    \item Usage for evaluating or scoring individuals, such as for employment, education, or credit
    \item Applying the model for critical automatic decisions, generating factual content, creating reliable summaries, or generating predictions that must be correct
\end{itemize}

\textbf{Misuse}

Intentionally using the model for harm, violating human rights, or other kinds of malicious activities, is a misuse of this model. This includes:

\begin{itemize}
    \item Spam generation
    \item Disinformation and influence operations
    \item Disparagement and defamation
    \item Harassment and abuse
    \item Deception
    \item Unconsented impersonation and imitation
    \item Unconsented surveillance
    \item Generating content without attribution to the model, as specified in the \href{https://huggingface.co/spaces/bigscience/license}{RAIL License, Use Restrictions}\\
\end{itemize}

\textbf{Intended Users}

\textbf{Direct Users}

\begin{itemize}
    \item General Public
    \item Researchers
    \item Students
    \item Educators
    \item Engineers/developers
    \item Non-commercial entities
    \item Community advocates, including human and civil rights groups
\end{itemize}

\textbf{Indirect Users}

\begin{itemize}
    \item Users of derivatives created by Direct Users, such as those using software with an intended use
    \item Users of \href{https://huggingface.co/spaces/bigscience/license}{Derivatives of the Model, as described in the License}
\end{itemize}

\textbf{Others Affected (Parties Prenantes)}

\begin{itemize}
    \item People and groups referred to by the LLM
    \item People and groups exposed to outputs of, or decisions based on, the LLM
    \item People and groups whose original work is included in the LLM\\
\end{itemize}

\textbf{Risks and Limitations}
This section identifies foreseeable harms and misunderstandings.

\begin{itemize}
\item Model may:
\begin{itemize}
\item Over-represent some viewpoints and under-represent others
\item Contain stereotypes
\item Contain personal information
\item Generate:
\begin{itemize}
\item Hateful, abusive, or violent language
\item Discriminatory or prejudicial language
\item Content that may not be appropriate for all settings, including sexual content
\end{itemize}
\item Make errors, including producing incorrect information as if it were factual
\item Generate irrelevant or repetitive outputs
\item Induce users into attributing human traits to it, such as sentience or consciousness\\
\end{itemize}
\end{itemize}

\textbf{Evaluation}

This section describes the evaluation protocols and provides the results.

\textbf{Metrics}

This section describes the different ways performance is calculated and why. Includes:

\textbf{Metric}: Perplexity. \textbf{Why chosen}: Standard metric for quantifying model improvements during training.

\textbf{Metric:} Cross Entropy Loss. \textbf{Why chosen:} Standard objective for language models.

And multiple different metrics for specific tasks. (More evaluation metrics forthcoming upon completion of evaluation protocol.)\\

\textbf{Recommendations}

This section provides information on warnings and potential mitigations.

\begin{itemize}
    \item Indirect users should be made aware when the content they're working with is created by the LLM.
    \item Users should be aware of Risks and Limitations, and include an appropriate age disclaimer or blocking interface as necessary.
    \item Models trained or finetuned downstream of BLOOM LM should include an updated Model Card.
    \item Users of the model should provide mechanisms for those affected to provide feedback, such as an email address for comments.
\end{itemize}

\end{document}